\documentclass[mathleft,fleqn,%
]{an}
\usepackage{graphicx}
\usepackage[varg]{txfonts}
\usepackage{natbib}
\bibpunct{(}{)}{;}{a}{}{,}
\def\teff{T_{\rm eff}}
\def\feh{\rm[Fe/H]}
\def\logg{\log\,g}

\def\msun{\rm M_{\odot}}

\def\dnu{\Delta\nu}
\def\num{\nu_\mathrm{max}}

\begin{document}

\Pagespan{1}{}
\Yearpublication{2014}%
\Yearsubmission{2014}%
\Month{0}%
\Volume{999}%
\Issue{0}%
\DOI{asna.201400000}%

\title{Asteroseismic age determination for dwarfs and giants}

\author{V. Silva Aguirre\inst{1}\fnmsep\thanks{Corresponding author:
        {victor@phys.au.dk}}
\and  A.\,M. Serenelli\inst{2}
}
\titlerunning{Asteroseismic ages of dwarfs and giants}
\authorrunning{Silva Aguirre \& Serenelli}
\institute{
Stellar Astrophysics Centre, Department of Physics and Astronomy, Aarhus University, Ny Munkegade 120, DK-8000 Aarhus C, Denmark.
\and 
Instituto de Ciencias del Espacio (ICE-CSIC/IEEC), Campus UAB, Carrer de Can Magrans, s/n, 08193 Cerdanyola del Valles, Spain}

\received{XXXX}
\accepted{XXXX}
\publonline{XXXX}

\keywords{stars: fundamental parameters -- stars: oscillations}

\abstract{%
Asteroseismology can make a substantial contribution to our understanding of the formation history and evolution of our Galaxy by providing precisely determined stellar properties for thousands of stars in different regions of the Milky Way. We present here the different sets of observables used in determining asteroseismic stellar properties, the typical level of precision obtained, the current status of results for ages of dwarfs and giants and the improvements than can be expected in the near future in the context of Galactic archaeology.}

\maketitle
\section{Introduction}\label{s:int}
The wealth of asteroseismic data from the CoRoT \citep{Michel:2008wx} and {\it Kepler} \citep{Gilliland:2010bb} missions has produced an authentic revolution in the field of stellar astrophysics. Thanks to the detection of oscillating modes in thousands of stars across the HRD, stellar properties can now be determined for large samples of targets with an unprecedented level of precision. This opens the exciting possibility of accurately characterising stellar populations in different regions of the Milky Way to constraint the history of formation and evolution of our Galaxy \citep[e.g.,][]{Miglio:2013hh,Casagrande:2014bd}. Of particular importance in this endeavour are precise age determinations for large cohorts of dwarfs and giants \citep{Chaplin:2014jf,Casagrande:2015je}, which can further help the usual kinematic and chemical dissection of the Galactic disc which has so far driven most of the comparisons between observations and simulations of our Galaxy \citep[e.g.,][just to name a few]{Edvardsson:1993uk,Schonrich:2009ee,Adibekyan:2012kr,Haywood:2013gw}.

We review the methods for determining ages of dwarfs and giants using different asteroseismic datasets and discuss their expected level of precision. All results presented here are based on two of the most sophisticated algorithms currently available for asteroseismic determination of stellar properties: the BAyesian STellar Algorithm \citep[BASTA,][]{SilvaAguirre:2015gi} and the Bellaterra Stellar Properties Pipeline \citep[][]{Serenelli:2013fz}. Briefly, both methods consist of a Bayesian approach including priors and appropriate weights to account for the volume space of the pre-computed grids of models used to construct the probability distribution functions. Different combinations of input observables can be included when determining stellar properties, which we discuss in detail below.
\section{Asteroseismic observables}\label{s:ast}
Detection of pulsation modes in dwarfs and red giants depends on the properties of convection in the outer stellar layers \citep[see,][and references therein]{Chaplin:2013gz}. Thus, detectability and data quality of asteroseismic properties for a certain length of observations is linked to the position of the star in the HRD as well as its intrinsic magnitude. We assume in the following a that determination of atmospheric parameters $\teff$ and $\feh$ is available for the stars in question.
\subsection{The bare minimum}\label{ss:scal}
A positive detection of oscillations in dwarfs or giants implies the appearance of a gaussian-shaped excess power in the Fourier transform of the time-series. This feature can be characterised by two so-called {\it global} asteroseismic observables: the frequency of maximum power $\num$ and the average large frequency separation $\langle\dnu\rangle$. The latter is the separation between modes of consecutive radial order and same angular degree, and is a measure of the travel time of the wave across the stellar interior. These quantities are related to the surface gravity and mean stellar density \citep[][]{Ulrich:1986ge,Brown:1991cv}, and thus form the basis of the asteroseismic scaling relations:
\begin{equation}\label{eqn:sca_mass} 
\frac{M}{\msun} \simeq \left(\frac{\num}{\nu_{\mathrm{max},\odot}}\right)^{3} \left(\frac{\langle\dnu\rangle}{\langle\dnu_\odot\rangle}\right)^{-4}\left(\frac{\teff}{T_{\mathrm{eff},\odot}}\right)^{3/2}\,, 
\end{equation}
\begin{equation}\label{eqn:sca_rad} 
\frac{R}{{\rm R_\odot}} \simeq \left(\frac{\num}{\nu_{\mathrm{max},\odot}}\right) \left(\frac{\langle\dnu\rangle}{\langle\dnu_\odot\rangle}\right)^{-2}\left(\frac{\teff}{T_{\mathrm{eff},\odot}}\right)^{1/2}\,, 
\end{equation}
where $\langle\dnu_\odot\rangle$, $\nu_{\mathrm{max},\odot}$ and $T_{\mathrm{eff},\odot}$ are the solar values. The global seismic parameters are what we call {\it the bare minimum}, that is the minimum amount of information that can be extracted from the time-series analysis given a positive detection. It is clear from Eqs.~\ref{eqn:sca_mass} and~\ref{eqn:sca_rad} that when these data are available, a direct estimation of the stellar mass and radius can be obtained \citep[e.g.,][]{Stello:2008gt,SilvaAguirre:2011es}. These scaling relations seem to hold well for radius determinations \citep[e.g.,][]{North:2007hl,Huber:2012iv,SilvaAguirre:2012du,White:2013bu}, while confirmation of masses is still under way using clusters and binaries.

\begin{figure}
\includegraphics[width=\linewidth]{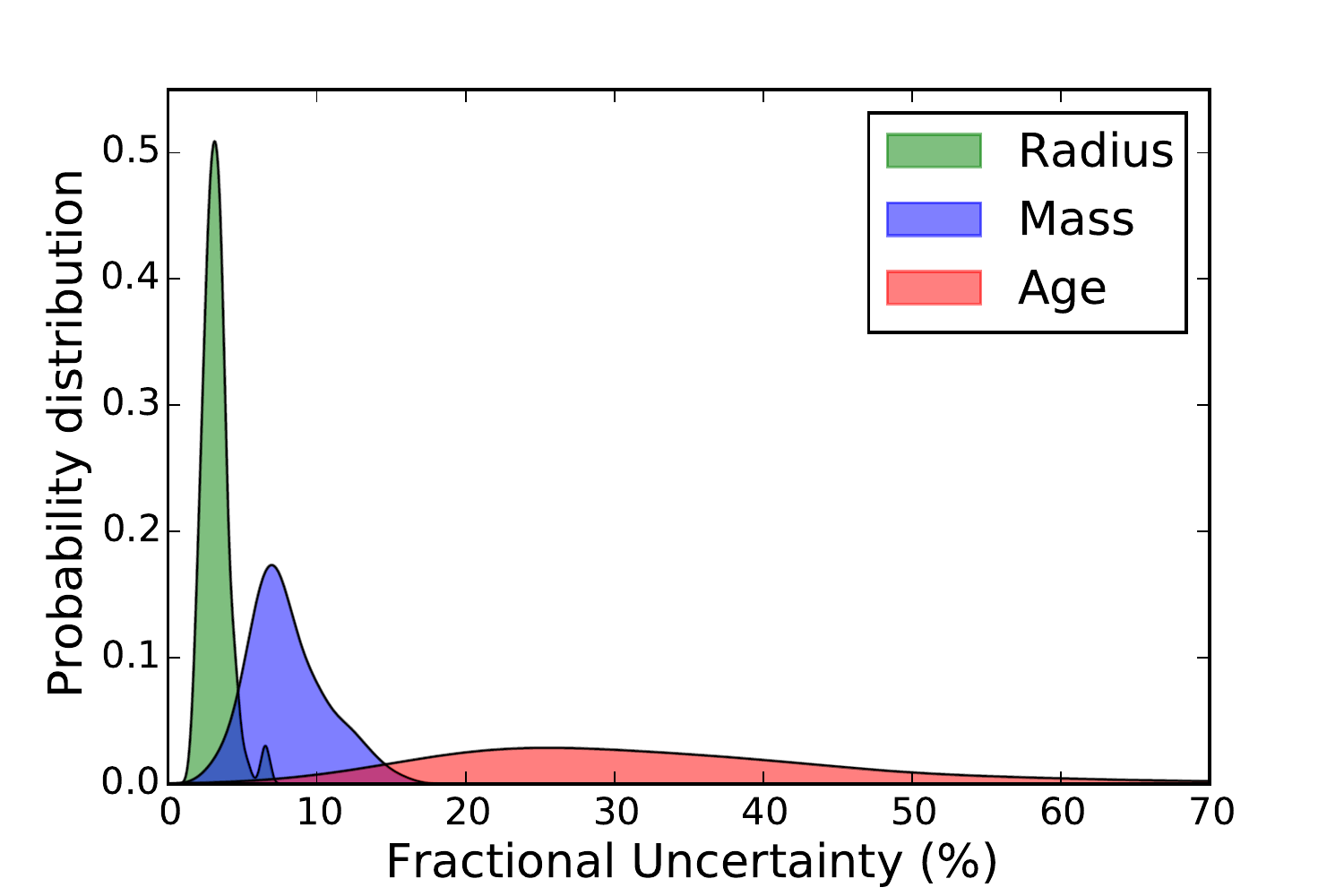}
\caption{Fractional uncertainties in stellar properties determined using Bayesian schemes and the global seismic parameters. Results for 87 dwarfs and subgiants analysed by \citet{Chaplin:2014jf}.}
\label{chaplin}
\end{figure}

In order to determine ages for stars using the global seismic parameters, BASTA and BeSPP compare these quantities and the atmospheric parameters to those predicted by grids of models. Figure~\ref{chaplin} shows the results obtained for the sample of 87 dwarfs and subgiants from \citet{Chaplin:2014jf} where spectroscopic $\teff$  and $\feh$ are available, resulting in median uncertainties of $\sim$2.2\%, $\sim$5.5\%, and $\sim$25\% in radius, mass, and age respectively (see Fig.~\ref{chaplin}). This level of precision is a factor of two better than half the targets of the Geneva-Copenhagen Survey \citep[see e.g., Fig.~16 in][]{2004A&A...418..989N}.

\begin{figure}
\includegraphics[width=\linewidth]{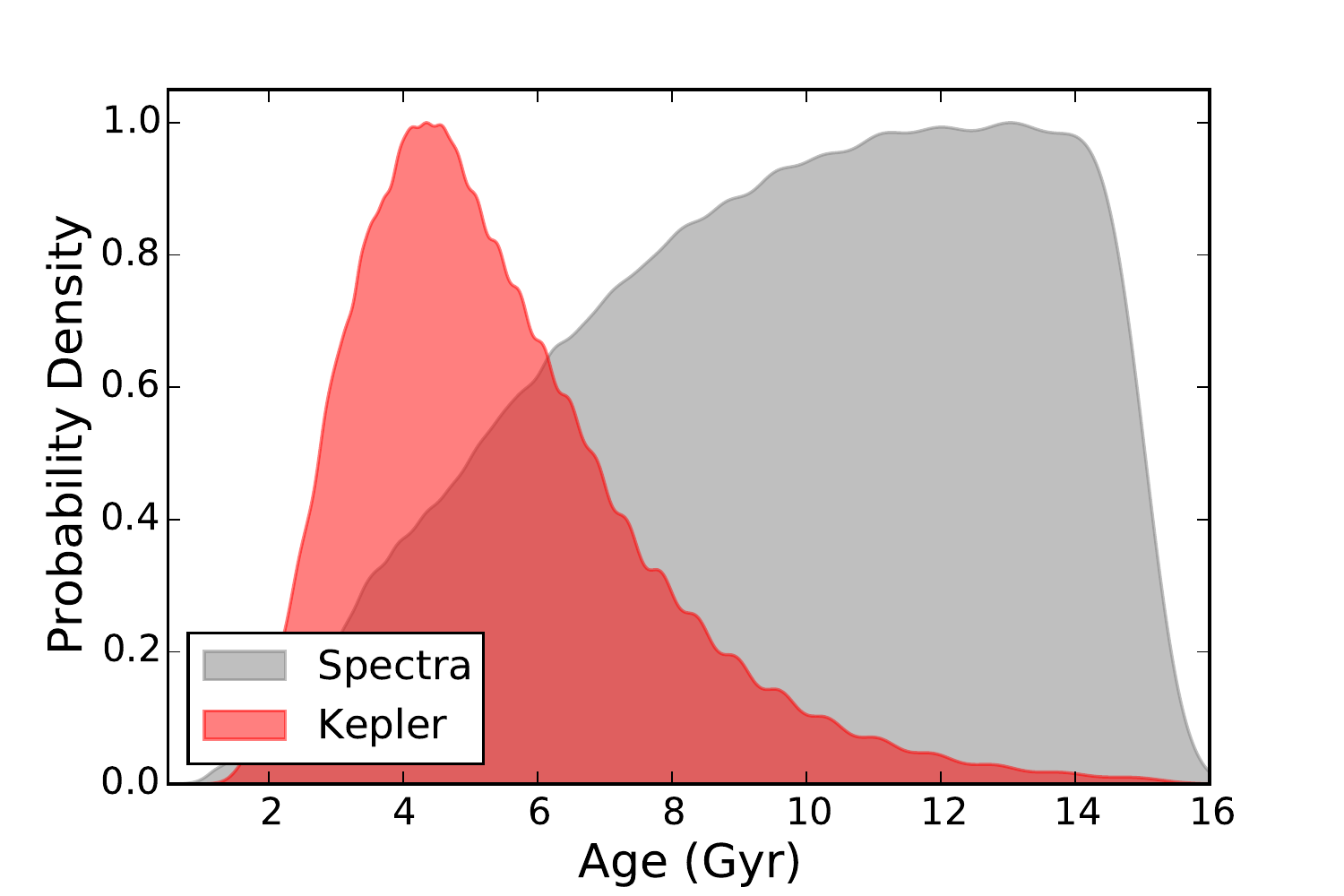}
\caption{Probability density function of age for a red giant star determined using a Bayesian scheme and different set of input: spectroscopic parameters $\teff$, $\logg$, and $\feh$ from the APOGEE survey (labelled Spectra) or global seismic parameters $\num$, $\langle\dnu\rangle$, complemented with $\teff$, and $\feh$ also from APOGEE (labelled Kepler).}
\label{spectra}
\end{figure}

In the case of red giants, the use of asteroseismology greatly improves the age determination as compared to classical isochrone placement.  Figure~\ref{spectra} shows the probability distribution function of age obtained using a Bayesian scheme and either the spectroscopic parameters $\teff$, $\logg$, and $\feh$ as input (labelled Spectra) or the global seismic observables complemented with $\teff$ and $\feh$ (labelled {\it Kepler}). It is currently not possible to obtain an age determination for giants based on atmospheric information only at the level required to support galactic studies (but see \citet{2015arXiv151108203M,2015arXiv151108204N} for promising developments on this issue). In stark contrast, ages based on asteroseismology achieve the required level of precision thanks to their sensitivity to the stellar interior. The first cohort of robustly determined ages for $\sim$1,000 red giant stars has recently been published by \citet{Casagrande:2014bd,Casagrande:2015je}. The level of precision achieved is of the order of $\sim$2.4\% and $\sim$6.0\% in radius and mass, respectively, while the median age uncertainty ranges between $\sim$20\% to $\sim$30\% depending on the evolutionary phase (see Section~\ref{ss:dp} below).
\subsection{Improvements for dwarfs and subgiants: individual frequencies}\label{ss:rats}
When the signal-to-noise ratios in the observations are high enough, it is possible to isolate the individual peaks in the power spectrum and perform what is known as "boutique" modelling of the targets \citep[e.g.,][]{2010ApJ...723.1583M,SilvaAguirre:2013in,Lebreton:2014gf}. In these cases, instead of fitting for the global asteroseismic parameters the Bayesian schemes aim at reproducing either the individual frequencies of oscillations or combinations of them. The first approach usually relies on an empirical surface correction to account for incomplete modelling of the outer stellar layers in 1-D hydrostatic codes \citep[e.g.,][]{Kjeldsen:2008kw}, while the latter suppresses the influence of these layers by building frequency rations \citep{Roxburgh:2003bb}.

\begin{figure}
\includegraphics[width=\linewidth]{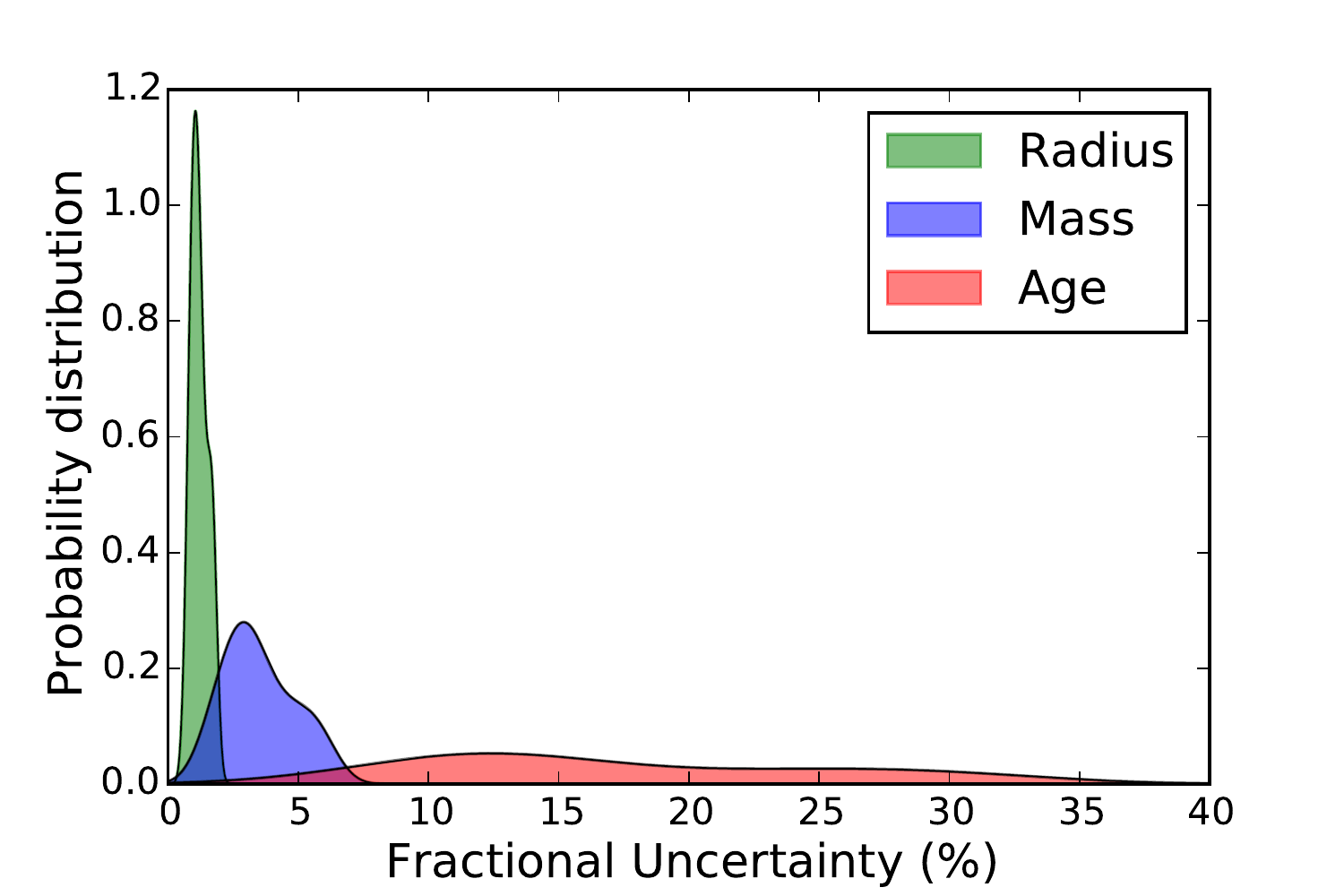}
\caption{Fractional uncertainties in stellar properties determined using Bayesian schemes and asteroseismic frequency ratios. Results for 33 exoplanet-host stars analysed by \citet{SilvaAguirre:2015gi}.}
\label{kages}
\end{figure}

Recently \citet{Lebreton:2014gf} showed that the most precise asterosesimic ages for dwarfs are those obtained using the frequency ratios as fitting observables as they are sensitive to the innermost layers of the star. The first homogeneous analysis of the 33 highest SNR {\it Kepler} exoplanet-host stars was made by \citet{SilvaAguirre:2015gi} using BASTA to reproduce these ratios, and the results are shown in Fig.~\ref{kages}. The median uncertainties in radius, mass, and age are of $\sim$1.1\%, $\sim$3.3\%, and $\sim$14\% respectively, almost a factor of two better than those obtained with the global seismic parameters for these type of stars.

The number of targets where this type of analysis is possible will increase to close to 100 dwarfs thanks to the upcoming {\it Kepler} LEGACY sample (Lund et al., in preparation; Silva Aguirre et al., in preparation). Although still far too low for Galactic studies, this sample will be the best characterised asteroseismic set available for the near future, and by virtue of being well distributed in the HRD its comprising stars will serve as benchmark for calibrating the properties of other stars where asteriseismic data are not available.

In the case of subgiant stars, their rapid core contraction after the main-sequence phase results in coupling between the pure acoustic and pure gravity modes cavities. Thus, non-radial frequencies of oscillations can present a mixed character behaving as g-modes in the stellar core and p-modes in the outer layers \citep[e.g.,][]{Aizenman:1977wh}. This behaviour results in a deviation from the asymptotic behaviour of pure p-modes, and the magnitude of that deviation can provide strong constraints on the core properties (and thus age) of subgiant stars. Initial investigations have found correlations between this mixed-mode character and stellar mass \citep[e.g.,][]{Deheuvels:2011fn,Benomar:2012kv}, but more studies are needed to properly characterise the sensitivity of these type of pulsations to stellar structure and the systematics involved in determining stellar properties by using them as the observables to fit.
\subsection{Improvements for red giants: period spacing}\label{ss:dp}
As was mentioned previously, mixed-modes carry information about the stellar core of stars that would otherwise be inaccessible to us. Of particular interests is the possibility of distinguishing between first ascent (RGB) and clump red giants, two classes of stars which occupy {\bf almost} the same region in the HRD thus making determination of their stellar properties by isochrone placement extremely challenging. Asteroseismology offers a window towards the structural region where these type of stars exhibit differences: while RGB stars have a radiative helium core surrounded by a hydrogen-burning shell, clump stars burn helium in their convective cores.

Detection of non-radial mixed modes by the CoRoT satellite \citep[][]{DeRidder:2009cd} opened the possibility of extracting information from the mixed-modes in red giants. While p-modes are equally separated in frequency by approximately the average large frequency separation $\langle\dnu\rangle$, g-modes are equally separated in period with a characteristic spacing dependent on the convective properties of the acoustic cavity. In other words, two red giants of very similar interior structure but one having a convective instead of a radiative core will show a different value of this period spacing \citep[see,][for a detailed explanation]{2014aste.book..194C}. Observational evidence came from the {\it Kepler} and CoRoT satellites, where \citet{Bedding:2011il} and \citet{Mosser:2011kg} showed that RGB and clump stars form two distinct sequences when comparing their measured mixed-modes period spacing.

\begin{figure}
\includegraphics[width=\linewidth]{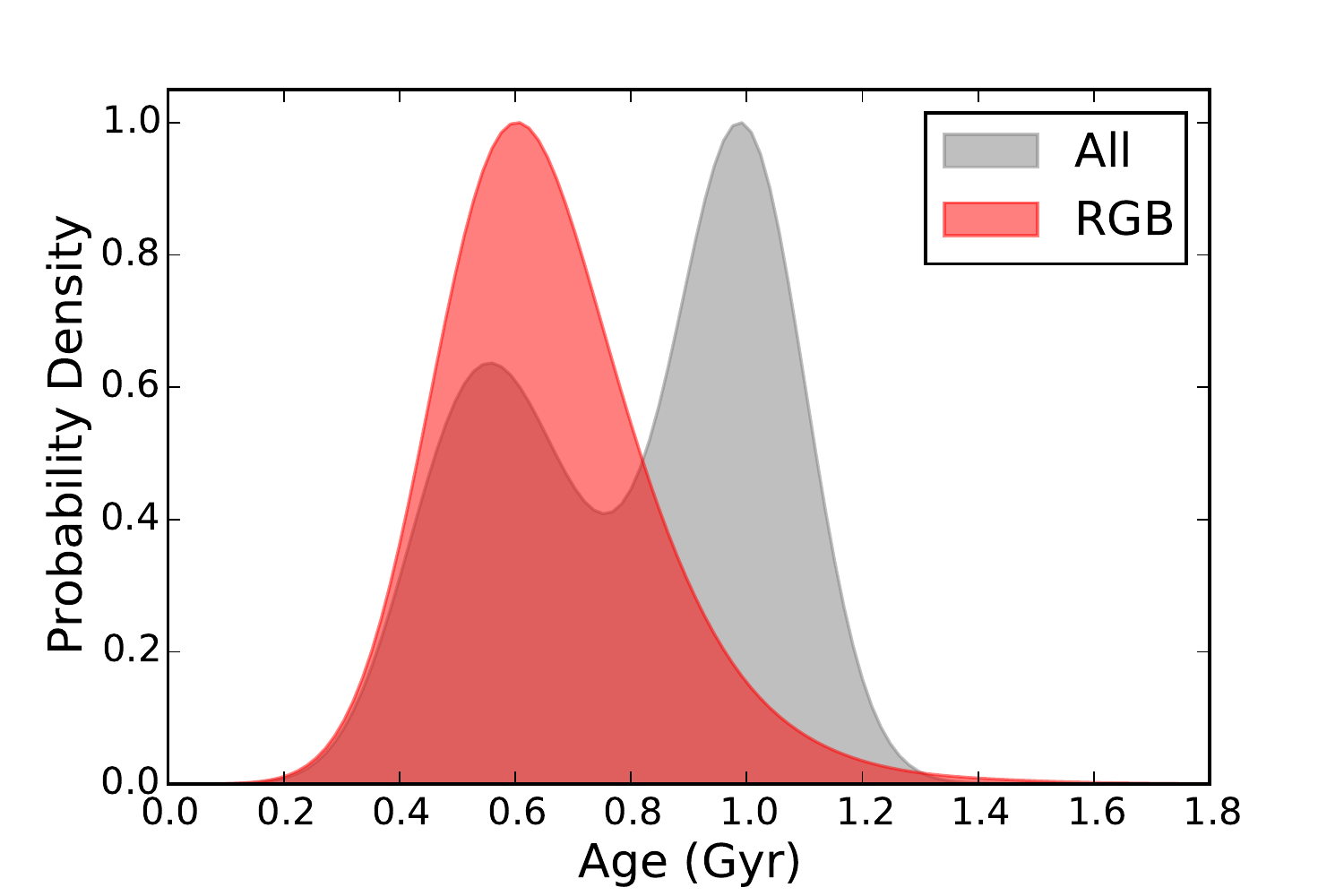}
\caption{Probability density function of age for a red giant star determined using a Bayesian scheme and the global seismic parameters $\num$, $\langle\dnu\rangle$, complemented with $\teff$, and $\feh$. Grey are depicts the age distribution when no information is available about the evolutionary stage (labelled All) while inclusion if this information as a Bayesian prior is plotted in red (labelled RGB).}
\label{giants}
\end{figure}

Figure~\ref{giants} shows the age determination for a {\it Kepler} target when information on the evolutionary phase is available. Implemented as a Bayesian prior, knowing that the star belongs to the RGB sequence obviously favours one of the peaks in the distribution and further constrains the age determination. A similar result was found by \citet{Casagrande:2015je}, where the authors showed that ages with uncertainties of $\sim$10-20\% where obtained for stars with a conclusive RGB identification from their period spacing. Uncertainties are slightly larger (up to $\sim$30\%) in ages for clump stars due to the unconstrained efficiency of mass-loss close to the RGB tip. Work is in progress to better determine its impact using asteroseismology of open clusters giants \citep[e.g.,][]{Miglio:2012dm}. 

\section{Conclusion and outlook}\label{s:con}
Asteroseismology is starting to deliver precise sets of stellar parameters for large cohorts of stars in different regions of the Milky Way. These data sets promise to become the new benchmarks for comparison of chemodynamical models going beyond the local volume covered by solar neighbourhood samples such as the Geneva-Copenhagen survey. We have described the methods to determine stellar properties using asteroseismology, in particular ages, based on different sets of inputs depending on data quality and availability. Ages of dwarfs, giants, and subgiants can currently be determined to a level of $\sim$20-30\% while further improvements can be made when individual frequencies or evolutionary classifications are available. The possibilities for conducting Galactic studies using these technique are immense considering the lines of sights currently being observed by the K2 mission \citep[][]{2015ApJ...809L...3S}, and the upcoming all-sky survey from the TESS satellite \citep[][]{Ricker:2015ie}.
\acknowledgements
Funding for the Stellar Astrophysics Centre is provided by The Danish National Research Foundation (Grant agreement no.: DNRF106). The research is supported by the ASTERISK project (ASTERoseismic Investigations with SONG and Kepler) funded by the European Research Council (Grant agreement no.: 267864). V.S.A. acknowledges support from VILLUM FONDEN (research grant 10118). A.M.S acknowledges support from the grants ESP-2013-41268-R (MINECO) and 2014SGR-1458.  
\bibliographystyle{an}
\bibliography{Badhonnef}
\end{document}